**Corresponding author**

# Hyosoung Cha

Cancer Data Center, National Cancer Center, 323 Ilsan-ro, Ilsandong-gu, Goyang 10408, Korea

E-mail: kkido@ncc.re.kr

# A Determination Scheme for Quasi-Identifiers Using Uniqueness and Influence for De-Identification of Clinical Data

Jipmin Jung(M.S.) [a], Phillip Park(M.S.) [a], Jaedong Lee(M.S.) [b], Hyein Lee(B.S.) [a], Geonkook Lee(M.D.,Ph.D.) [c], Hyosoung Cha(Ph.D.) [a].

[a] Cancer Data Center, National Cancer Center, Goyang, Korea

[b] Department of Information Technology, National Cancer Center, Goyang, Korea

[c] Office of Public Relations and Collaboration, National Cancer Center, Goyang, Korea

## Abstract

Objectives; The accumulation and usefulness of clinical data have increased with IT development. While using clinical data that needs to be identifiable to obtain meaningful information, it is essential to ensure that data is de-identified and unnecessary clinical information is minimized to protect personal information. This process requires criteria and an appropriate method as there are clear identifiers as well as quasi-identifiers that are not readily identifiable.

Methods; To formulate such a method, first, primary quasi-identifiers were selected by classifying information in 20 clinical personal information database tables into Direct-Identifier (DID), Quasi-Identifier (QI), Sensitive Attribute (SA), and Non-Sensitive Attribute (NSA) according to its type. Secondary QIs were then selected by assessing the risk for outliers by measuring uniqueness values of the selected data and scoring re-identification by calculating equivalence class of the influence on other data on QI removal. Third, the risk of re-identification of data users was numeralized and classified. Lastly, the final QI according to user class was determined by comparing the calculated re-identification scores to the threshold values of user classes.

Results; Eventually, final QIs ranging from a minimum of 18 to a maximum of 28 were selected by making an assumption about user classes and using it as criteria.

Conclusions; The QI selection method presented by the current investigators can be used by researchers at the final checkup stage before they de-identify the selected QIs. Therefore, clinical data users can securely and efficiently use clinical data containing personal information by objectively selecting QIs using the method proposed in the present study.

## Keywords

health information management; electronic health records; medical information; clinical data; data de-identification;

# 1. Introduction

The big data industry has made it possible to perform analyses in various areas; however, the industry is now able to provide individualized services to information users by converging and utilizing each individual's data. Therefore, although the value of data utilization has increased, the risk of loss of personal information has increased as well.[1, 2]

Medical data include various medical records and results based on the actions taken with regard to the treatment of an individual at a hospital such as basic information about the patient, pathology, admission, discharge, and surgery information. In addition, medical data also include sensitive personal information such as disease codes and surgery history. Therefore, de-identification must be performed prior to using medical data for non-medical purposes.[3]

De-identification is a process used to make the subject of the information unidentifiable. Personally identifiable information such as name and resident registration number are included in information that can identify an individual.[4]

Medical data can be classified in the context of de-identification into Direct-Identifier (DID), Quasi-Identifier (QI), Sensitive Attribute (SA), and Non-Sensitive Attribute (NSA). DID refers to data that enable direct identification of a target individual such as name, social security number, and e-mail address. In the United States, Health Insurance Portability and Accountability Act (HIPAA) Privacy rules regulate the utilization of medical information and selection of DID type based on laws related to the personal information protection.[5, 6] Because such DIDs reveal identity without any other additional information, de-identification measures are taken to prevent identification by either removing the value corresponding to the selected DID or by applying predetermined rules.[7]

QI is defined as the information that can identify an individual when combined with other categories of information even though they are not a DID. It is also important personal information that requires the same level of processing as a DID. A large amount of data loss will occur if the information classified as QI is de-identified, and the data that can be used for actual analysis will be greatly reduced. Accordingly, selecting QIs appropriately and providing information that is actually needed for data analysis are important issues associated with de-identification.[8]

Because no exact criteria have been defined for de-identification until now, QIs are selected by subjective judgments stemming from the experience of the person in charge. In addition, QI selection is inconsistent because it is de-identified based on different criteria. When such data are collated, there is a possibility of the identity of the information subject being revealed by privacy attacks such as linkage attacks.[9] On the other hand, the QIs selected based on the guidelines for de-identification of personal information comply with criteria for the prevention of information re-identification. Therefore, a QI selection method needs to be established with consideration to using medical data, distribution range, and QI characteristics given that data utility will be reduced owing to most of the data being suppressed if all defined example items are selected as QIs and de-identified.[10, 11]

The QI related research mostly proposed so far has involved comparing and analyzing target data, and based on the outcome, data on new de-identification targets is suggested. Research conducted until now, however, also has limitations in that there was obscure standard, and de-identification methods judged to be the most appropriate were subjectively suggested.

Therefore, we attempt to suggest a methodology that can objectively specify QIs within the scope of clinical data. The proposed methodology can minimize the danger of information leakage when used for research by identifying connections in data based on the risk of exposing information when data is collated and minimize data loss through de-identification processes.

## 1.1 Background

The information that can identify an individual by itself should be eliminated. In addition, risks due to combining with other information should be minimized. It means that de-identification may be needed to accomplish the analysis goal in the case that obtaining consent for analyzing retained personal information is difficult. Considering the risk of information re-identification, management and post management of de-identification should be thorough.[5, 6]

Fundamentally, de-identification technologies assume the risk of re-identification. The level of risk is considered to differ depending on the de-identification method applied. Various elements are included, but the level of specific data linked to a particular individual is one of the most basic elements. From this perspective, the risk of personal information infringed upon due to the identification of applicable information varies depending on whether the data are linked to a specific individual, whether there is a potential possibility of data linking to a specific individual, or whether the data can possibly be linked to a loosely defined group of people instead of a specific individual.[7, 12]

A comparison of de-identification related studies in the last five years reveals significant progress in the de-identification research on clinical information. Studies so far on various de-identification methods have primarily focused on avoiding linkage attacks.[1, 13-16] They have mostly researched methods to minimize the identifiability of individuals from combining pieces of information by strengthening de-identification methods. Furthermore, such research methods have been focused on retrospective studies that use data from clinical information.[17-26] The reason is that it is easy to be exposed to the risk of re-identification in the process of reusing data.[2] Although the risk of data re-identification can be reduced with such research methods, the possibility of reducing data utility also exists.[7] Consequently, research should be conducted on maintaining the value of data through studies on procedures for the selection of QIs, which are the targets of de-identification. Even though there have been studies dealing with QIs, they simply treat QIs as targets for de-identification, and research on selecting data as QIs based on specific criteria has not been conducted. Accordingly, taking this research background into consideration, the present study aims to investigate data utility and minimization of re-identification risk by selecting QIs using specific criteria only from clinical data for retrospective studies.

## 2. Methods

### 2.1 De-identification analysis targets

De-identification analysis was performed on Electronic Medical Record (EMR), which was built in National Cancer Center in Korea, as source data. The EMR is composed of 6,761 tables containing four types of clinical data: clinical code, prescription and results, patient information, and registration. Experiments were conducted on 17 tables that did have personal information. Additionally, even though death-related information is not personal information according to the act for personal information protection, 3 tables containing death-related information were added since it is included in the review as information to be protected.

### 2.2 Final QI selection process

We use the scores of the user who analyzes the data for research such as retrospective study and the institution with which the user is affiliated as the criterion for selecting QIs to identify the objective balance between re-identification risk and data utility. The items known to be elements that can identify an individual were first selected as QIs by combining them with other information according to the HIPAA rules.[5] The calculation of uniqueness and influence values is described at step 2. The final QIs were selected for the digitized columns based on the QI selection criteria established using the scores for the user and institution. Figure 1 shows the process carried out to select the final QIs.

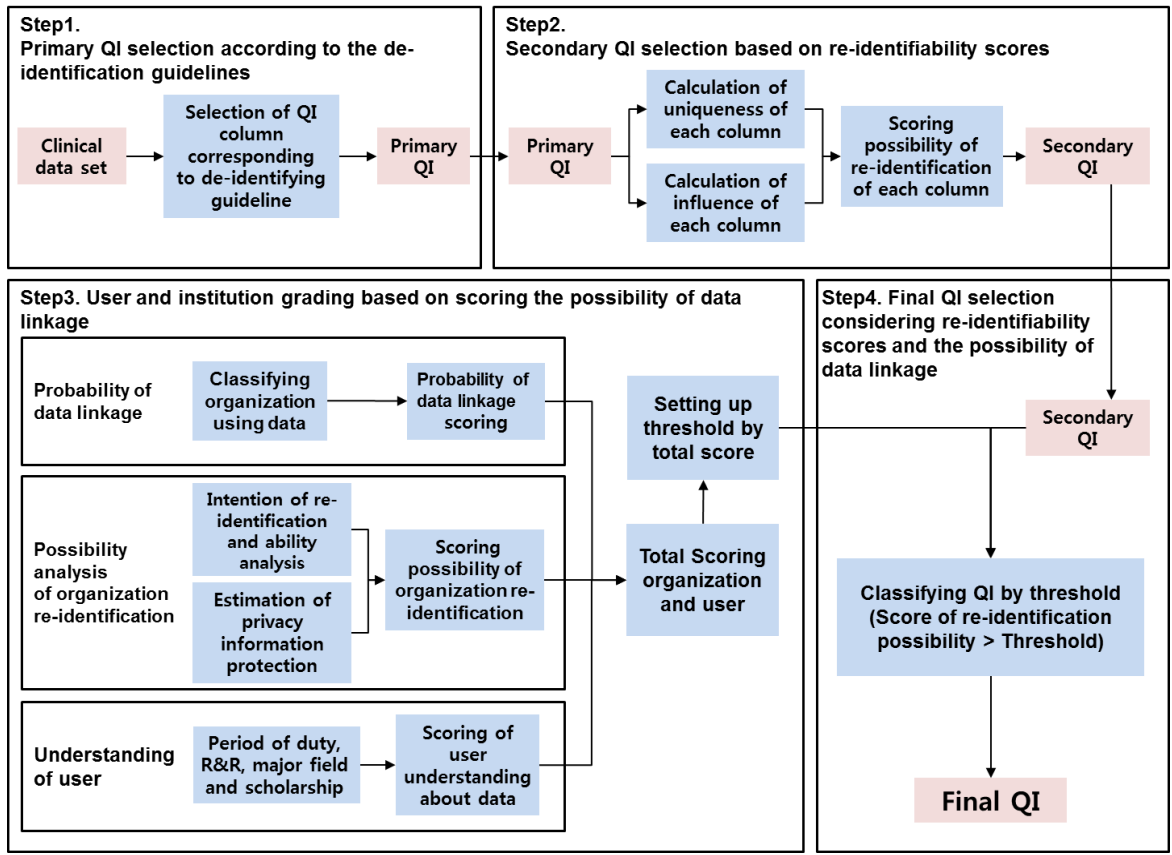

Figure 1. Process of QI decision by possibility of data linkage, analysis of re-identification possibility in organization, understanding of user, and possibility of re-identification.

## Step 1. Primary QI selection according to the de-identification guidelines

Selection of QIs aims not only to safely use data but also to obtain data that must be de-identified to use in research. Accordingly, de-identification of QIs is subjected to different criteria depending on the purpose of research, unlike DIDs that require compulsory deletion for de-identification. In addition, the sensitive data included in clinical data must also be deleted in principle if they are irrelevant to the purpose of data use. Therefore, the first step in the de-identification process is classifying data columns into DID, QI, SA, and NSA.[6] The data column classification process is indicated in Figure 2. If the information is not a DID and a specific individual cannot be identified using the information but can be easily identified by combining it with other information such as personal, physical, and credit characteristics, the information is classified as a QI. If an individual is identified using information that is not a DID or QI, sensitive information that can inflict ethical, financial, and reputational harm upon an individual is classified into an SA, and other general information is classified into an NSA.

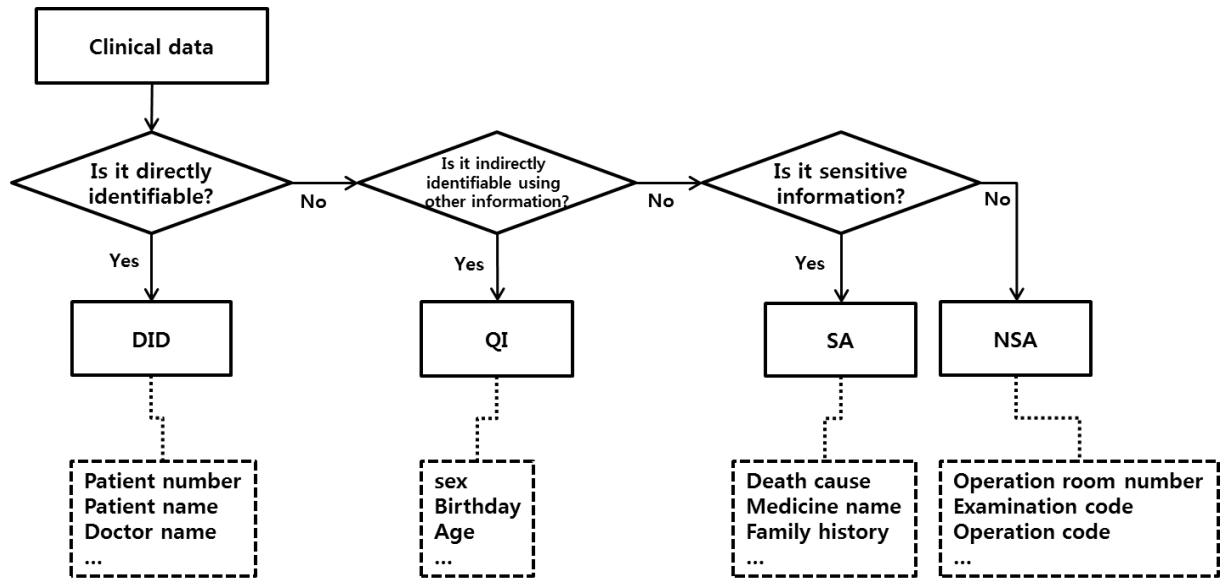

Figure 2. Process of classification by data column for de-identification. DID: direct-identification; QI: quasi-identification; SA: sensitive attribute; NSA: non-sensitive attribute.

## Step 2. Secondary QI selection based on re-identifiability scores

### A. Calculation rule of uniqueness

$$Uniqueness\ value = \frac{number\ of\ unique\ values\ in\ column}{total\ number\ of\ different\ values\ in\ column}$$

| | Weight | Age | Gender | Zipcode |
|---|---|---|---|---|
| | 72 | 45 | M | 75145 |
| | 72 | 45 | M | 75145 |
| | 58 | 21 | M | 47853 |
| | 45 | 21 | F | 47853 |
| | 45 | 64 | F | 47853 |
| Uniqueness | 1/5 | 1/5 | 0/5 | 0/5 |

### B. Calculation rule of influence

$$Influence\ value = 1 - \frac{N_E(T-C_i)}{N_E(T)},\ i = 1, 2, \ldots, n$$

Where $N_E$ is number of equivalence class,
$T = \bigcup_{i=1}^{n} C_i$, is number of column
and $C_i$ is the $i_{th}$ column

| Weight | Age | Gender | Zipcode |
|---|---|---|---|
| 72 | 45 | M | 75145 |
| 72 | 45 | M | 75145 |
| 58 | 21 | M | 47853 |
| 45 | 21 | F | 47853 |
| 45 | 64 | F | 47853 |

| Influence value of weight column | Influence value of weight column | Influence value of weight column | Influence value of weight column |
|---|---|---|---|
| (1-4/4)=0 | (1-3/4)=0.25 | (1-4/4)=0 | (1-4/4)=0 |

| Weight | Age | Gender | Zipcode | Weight | Age | Gender | Zipcode | Weight | Age | Gender | Zipcode | Weight | Age | Gender | Zipcode |
|---|---|---|---|---|---|---|---|---|---|---|---|---|---|---|---|
| 72 | 45 | M | 75145 | 72 | 45 | M | 75145 | 72 | 45 | M | 75145 | 72 | 45 | M | 75145 |
| 72 | 45 | M | 75145 | 72 | 45 | M | 75145 | 72 | 45 | M | 75145 | 72 | 45 | M | 75145 |
| 58 | 21 | M | 47853 | 58 | 21 | M | 47853 | 58 | 21 | M | 47853 | 58 | 21 | M | 47853 |
| 45 | 21 | F | 47853 | 45 | 21 | F | 47853 | 45 | 21 | F | 47853 | 45 | 21 | F | 47853 |
| 45 | 64 | F | 47853 | 45 | 64 | F | 47853 | 45 | 64 | F | 47853 | 45 | 64 | F | 47853 |

Figure 3. Example of calculation rules of uniqueness (A) and influence (B).

The QIs are selected based on identifiability scores determined by identifying the characteristics of the data. Uniqueness was calculated for each column, since a column with a large number of unique values can be considered to have a high possibility of re-identification (Figure 3A).[27] For example, the uniqueness of a column means the calculated ratio of unique records, i.e., The ratio of a record with only one value to all the records in the column. Therefore, because 58 and 64 are unique values among the five values in each column in the cases of weight and age, respectively, they have a uniqueness value of 0.2 each. If the uniqueness value is 0, the information cannot identify a specific individual, implying that it does not have to be selected as a QI because of its low data risk. On the other hand, a column that has a non-zero uniqueness value implies that it has at least one distinct value. Such a column can be considered to have re-identification risk because the distinct value can be key to identifying a specific individual.

Next, the level of influence is measured based on changes in the number of equivalence classes (Figure 3B). $N_E$ denotes the number of equivalence classes in a data set, T is the sum of all columns in the table, and $C_i$ denotes the $i_{th}$ column. If the number of equivalence classes decreases sharply compared to the number of equivalence classes of the entire table when a specific column is excluded, then the level of influence of the specific column on the data is high. In Figure 3A, even though the uniqueness values of weight and age are the same, the high influence of age can be seen if the ratio of the number of equivalence classes in each column to that in the entire table is calculated. Because the number of equivalence classes of the entire table is four while it is three when the age column is excluded, the influence value of age is 0.25. A column with an influence value indicates that it can increase the risk of re-identification by increasing the number of equivalence classes due to its influence on other columns, even if its uniqueness value is 0. Therefore, it is selected as a QI, which has the possibility of identifying individuals by combining with other data.

Lastly, among the primary QIs, the QIs that have uniqueness and influence values calculated above are selected as the secondary QIs and scored as the sum of uniqueness and influence to use them as re-identifiability scores.

## Step 3. User and institution grading based on scoring the possibility of data linkage

The final QI selection reflects the risk of data re-identification through a survey of the institution and users using the data. First, scoring is carried out by assigning weights based on the characteristics of the institution to prevent identification of specific individuals with data QIs provided through the data institutions have by classifying institutions that use data. Second, scoring is performed after weightings are assigned to the possibility of re-identification by assessing it and the ability to protect the personal information of data users. Lastly, scoring is performed based on the data user's ability to protect personal information. Lastly, the re-identification of data is scored. Table 1 shows an example of indices that can be scored on the possibility of data linkage. Institutions and enterprises using clinical research data assign weights according to the ease of collecting data related to the applicable data and score. In the case of public institutions, gradings are carried out with differentiation so that institutions that engage significantly with clinical data such as the National Health Insurance Service, the Health Insurance Review & Assessment Service, and Public Hospitals are graded high; institutions that engage relatively lesser such as the Ministry of Health & Welfare and the Ministry of Food and Drug Safety are graded middle, and other institutions that engage the least are graded low. Among enterprises, hospitals, which have high relevance to clinical data, are graded high; pharmaceutical companies, which have relatively low relevance are graded middle; and other enterprises are graded low. To differentiate the grades determined above, scores of ten, five, and one are assigned to high, middle, and low grades, respectively.

Table 1. Probability example of data linkage

| Categorization | Institution | Probability | Grade |
|---|---|---|---|
| **Public Institution** | National Health Insurance Service, Health Insurance Review & Assessment Service, Public Hospital, etc. | High | High |
| | Ministry of Health & Welfare, Ministry of Food and Drug Safety, etc. | Middle | Mid |
| | Government Agency and so on | Low | Low |
| **General Enterprise** | Hospital | High | High |
| | Pharmaceutical Company | Middle | Mid |
| | Enterprise and so on | Low | Low |

Re-identifiability was divided into user such as researcher for retrospective study, and clinical data aspects to assess the re-identifiability of institutions. For the user aspect, users' intent and ability to re-identify clinical research data are measured. For the data aspect, the level of privacy information protection such as procedures to protect data itself are measured and scored. Table 2 is a summary of detailed indicators of re-identification intent of the user aspect and the possibility of external data linkage of the data aspect. For re-identification intent, the possible impact of re-identified data using Yes/No. For the possibility of external data linkage, the possibility of users linking external information to re-identify is assessed using Yes/No. The total score is obtained by summing the number of "Yes" responses with one point assigned for each response and the maximum score being four points as there are four indicators. The higher the score, the higher the intent and ability for re-identification.

Table 2. Intention of re-identification and ability analysis1

| Classification | Detailed Indicator | Evaluation |
|---|---|---|
| **Intention of re-identification** | Data frame that aims to shames individuals when data users or consumers re-identify the data | Yes/No |
| | Data frame that offers monetary benefits when data users or consumers re-identify the data | Yes/No |
| | Data users or requestors who do not communicate about the prohibition of re-identification and restriction of data provision to third party in data usage (provision) related contract | Yes/No |
| **Possibility of external data linkage** | Data that is combinable with the evaluation-target data available on the Internet, SNS, and data.go.kr (government open-data portal) | Yes/No |

Table 3 summarizes the detailed indicators used to assess the level of privacy information protection. It should be noted that the indicators are used to measure the risk for protection level, and the number of "No" assessments should be summed. Accordingly, the higher score the score, the lower is the protection level. The maximum score is six points.

Table 2. Estimation of privacy information protection

| Classification | Detailed Indicator | Evaluation |
|---|---|---|
| **Ability in privacy protection** | Receiving security service level agreement or providing security training to human resources who can access the data | Yes/No |
| | Establishing or operating management plan about keeping and treating data for users or requestors | Yes/No |
| | Data is transferred through a safe method that is physically and technologically protected | Yes/No |
| | It is used on a server or PC with both systems of intrusion blocking and intrusion prevention installed | Yes/No |
| | Managing the access authorization and access records of human resources that can access data | Yes/No |
| | Data users or requestors undergo periodic security check from security management division | Yes/No |

The scores for re-identification intent are digitized by adding assessment scores in Tables 2 and 3. The obtained score is a re-identifiability index and the higher the score, the higher the risk of re-identification , implying that data protection should be strengthened.

Users' understanding of data such as clinical data for research is measured separately for the relevant knowledge of data users and length of employment. For the relevant knowledge, users' understanding of data is assessed and scored by analyzing their role and responsibilities (R&R), major, and academic degree. For the items on length of employment, given the background knowledge on work, it is considered that the longer the length of employment, the higher the re-identification risk. Table 4 shows the classification of user's ability and explanation on detailed indicators. The relevant knowledge is assessed as Yes/No, and the length of employment is scored. One point is assigned to "Yes" on the detailed items related to relevant knowledge, with the maximum score being three points. Items related to the length of employment were scored by assigning zero point for less than three years of employment, three points for a period between more than three years and less than seven years, five points for a period between more than seven years and less than ten years, and seven points for ten years or more. Users' understanding of data was scored by summing the scores of relevant knowledge and length of employment. The higher the score, the higher the risk of user's re-identification, and the greater the need for strengthening data protection.

Table 3. User understanding about data

| Classification | Detailed Indicator | Evaluation |
| --- | --- | --- |
| **Relevant knowledge** | Data users or requestors have knowledge or relevant degrees indicating the ability to re-identify private information | Yes/No |
| | Data users or requestors can possess or obtain resources (money) that can be used to re-identify private information | Yes/No |
| | Data users or requestors can access other databases that can be linked for the re-identification of private information | Yes/No |
| **Working period** | Working period of data users or requestors is less than 3 years | 0 pts |
| | Working period of data users or requestors is 3 or more years but less than 7 years | 3 pts |
| | Working period of data users or requestors is 7 or more years but less than 10 years | 5 pts |
| | Working period of data users or requestors is 10 or more years | 7 pts |

The average of the scores of the possibility of data linkage, analysis of the re-identifiability of institutions, and users' understanding of data is obtained. The average value is entered as Average Score, and grade is calculated. The average scores are classified into "High" for seven points or higher, "Middle" for a score higher than four points but less than seven points, and "Low" for less than four points. For example, assuming ten points for the possibility of data linkage, five points for re-identifiability, and five points for users' understanding of data, the Average Score of institution and users is 6.67, which is classified as "Middle" (Table 5).

Table 4. Example of calculation for organization and user

| Probability of data linkage | Possibility analysis of organization re-identification | User understanding | Average score |
| --- | --- | --- | --- |
| 10* | 2** | 8*** | 6.67 (Middle) |

\* The score is obtained from the contents of Table 1, and the score ranges from a minimum of one point to a maximum of ten points.

\*\* The score is obtained from the contents of Tables 2 and 3, and the score ranges from a minimum of one point to a maximum of ten points.

\*\*\* The score is obtained from the contents of Table 4, and the score ranges from a minimum of one point to a maximum of ten points.

## Step 4 Final QI selection considering re-identifiability scores and the possibility of data linkage

To differentiate the selection criteria for the final QIs depending on the users using data and their affiliated institutions, the final QI selection is determined for the re-identifiability of scores calculated in Step 3 based on the calculated threshold according to the extracted grade. A threshold value of data identifiability of 0.25 is assigned for "High" grade, 0.5 for "Middle" grade, and 0.75 for "Low" grade respectively. Only the columns that scored higher re-identifiability than the applicable thresholds are selected as QIs. The lower the grade, the higher the threshold, which results in less number of QIs being selected.

## 3. Results

The clinical data set tables were classified into DIDs, QIs, SAs, and NSAs according to the existing HIPAA rules. Table 6 indicates the classification status of the 20 CRDW tables, which were classified into 108 DIDs, 137 QIs, 68 SAs, and 460 NSAs. The number of classified DIDs, QIs, SAs, and NSAs includes duplicate counts.

Table 6. Analysis results of clinical data set tables by National Cancer Center

| No | Description | DID | QI | SA | NSA |
|---|---|---|---|---|---|
| 1 | Patient master | 4 | 9 | 5 | 10 |
| 2 | Physical measurement information | 7 | 11 | 0 | 12 |
| 3 | Patient mortality information | 1 | 1 | 4 | 4 |
| 4 | Patient mortality date information | 1 | 1 | 0 | 4 |
| 5 | Patient mortality cause information | 1 | 0 | 4 | 4 |
| 6 | Therapeutic radiation therapy | 12 | 10 | 0 | 27 |
| 7 | Processing prescription details | 4 | 6 | 0 | 22 |
| 8 | Blood transfusion prescription details | 4 | 12 | 0 | 37 |
| 9 | Details of rehabilitation treatment | 6 | 8 | 0 | 27 |
| 10 | Reading results TA | 1 | 1 | 0 | 43 |
| 11 | Pathological reading results | 1 | 1 | 0 | 6 |
| 12 | Results of pathology readings | 10 | 6 | 3 | 21 |
| 13 | Surgical prescription details | 6 | 11 | 0 | 28 |
| 14 | Image function test results | 15 | 5 | 0 | 15 |
| 15 | Medication prescription details | 4 | 7 | 6 | 61 |
| 16 | Diagnostic test results | 3 | 3 | 4 | 30 |
| 17 | Test prescription details | 4 | 6 | 0 | 34 |
| 18 | Diagnostic information | 6 | 8 | 0 | 16 |
| 19 | Visit information | 10 | 3 | 0 | 22 |
| 20 | Early assessment of nursing (General adult) | 8 | 28 | 42 | 37 |

In the 108 columns classified into DID among the 20 tables, the DIDs associated with patients are patient names, town-level addresses, and patient identification numbers. In addition, the DIDs related to hospital personnel were doctor ID, doctor name, therapist ID, therapist name, anesthesiologist ID, anesthesiologist name, initial keyboarder, and final amender.

Of the 137 columns selected as QIs, the columns with uniqueness scores and influence were considered as re-identifiable columns, and the first batch of 64 QIs including duplicates were selected. Table 7 illustrates the method of making final QI selections for the applicable columns based on the threshold values graded using objective scores for the users and institutions mentioned above. First, re-identifiability was scored by summing the uniqueness scores and influence scores of each column. From the obtained scores, 28 columns were selected as QIs if the threshold was set at 0.25 and by taking columns that have re-identifiability score of 0.25 or higher as QIs in the case of "High" user and institution grading. A total of 20 columns were selected as QIs if the threshold was set at 0.5 and by taking columns with re-identifiability score of 0.5 or higher as QIs in the case of "High" user and institution grading. Lastly, 17 columns were selected as QIs if the threshold was set at 0.75 and by taking columns with re-identifiability score of 0.75 or higher as QIs in the case of "High" user and institution grading

Table 7. The list of selected columns in CRDW

| No | Table description | Column description | Uniqueness | Influence | Sum (Uniqueness +Influence) | QI (0.25+/ check) | QI (0.5+/ check) | QI (0.75+/ check) | No | Table description | Column description | Uniqueness | Influence | Sum (Uniqueness +Influence) | QI (0.25+/ check) | QI (0.5+/ check) | QI (0.75+/ check) |
|---|---|---|---|---|---|---|---|---|---|---|---|---|---|---|---|---|---|
| 6 | Therapeutic radiation therapy | Gender | 0 | 0.0008 | 0.0008 | | | | 13 | Surgical prescription details | Medical charge code | 0.0003 | 0 | 0.0003 | | | |
| | | Date of birth | 0.0000 | 0.6494 | 0.6494 | O | O | | | | Medical charge name(English) | 0.0003 | 0 | 0.0003 | | | |
| | | Age at the prescription | 0 | 0.0748 | 0.0748 | | | | | | Medical charge name(Korea) | 0.0003 | 0 | 0.0003 | | | |
| | | Therapeutic site code | 0 | 0.1401 | 0.1401 | | | | 14 | Image function test results | Gender | 0 | 0.2927 | 0.2927 | O | | |
| 7 | Processing prescription details | Gender | 0 | 1 | 1 | O | O | O | | | Date of birth | 0.0003 | 0.9994 | 0.9997 | O | O | O |
| | | Date of birth | 0.0000 | 0.9947 | 0.9947 | O | O | O | | | Age at the examination | 0 | 0.8670 | 0.8670 | O | O | O |
| | | Age at the prescription | 0 | 1 | 1 | O | O | O | 15 | Medication prescription details | Gender | 0 | 0.0663 | 0.0663 | | | |
| | | Prescription type code | 0 | 1 | 1 | O | O | O | | | Date of birth | 0.0001 | 0.9757 | 0.9758 | O | O | O |
| | | Prescription code | 0.0000 | 0.9997 | 0.9997 | O | O | O | | | Age at the examination | 0 | 0.4918 | 0.4918 | O | | |
| | | Medical charge code | 0.0000 | 0.9997 | 0.9997 | O | O | O | | | Prescription code | 0 | 0.0025 | 0.0025 | | | |
| 8 | Blood transfusion prescription details | Gender | 0 | 0.0135 | 0.0135 | | | | | | Medical charge code | 0 | 0.0195 | 0.0195 | | | |
| | | Date of birth | 0.0095 | 0.8194 | 0.8289 | O | O | O | | | Medical charge code(English) | 0 | 0.4085 | 0.4085 | O | | |
| | | Age at the prescription | 0 | 0.2026 | 0.2026 | | | | | | Medical charge code(Korea) | 0.0000 | 0.6785 | 0.6785 | O | O | |
| | | Prescription code | 0 | 0.0110 | 0.0110 | | | | 16 | Diagnostic examination results | Gender | 0 | 0.3070 | 0.3070 | O | | |
| | | Operation name | 0.0077 | 0.0189 | 0.0266 | | | | | | Date of birth | 0.0000 | 0.9994 | 0.9994 | O | O | O |
| | | Blood type | 0 | 0.0082 | 0.0082 | | | | | | Age at the examination | 0.0000 | 0.8680 | 0.8680 | O | O | O |
| 9 | Details of rehabilitation treatment | Gender | 0 | 0.0035 | 0.0035 | | | | 17 | Examination prescription details | Gender | 0.0009 | 0.0749 | 0.0758 | | | |
| | | Date of birth | 0.0027 | 0.8187 | 0.8214 | O | O | O | | | Date of birth | 0 | 0.0061 | 0.0061 | | | |
| | | Age at the prescription | 0 | 0.1348 | 0.1348 | | | | | | Age at the prescription | 0 | 0.0005 | 0.0005 | | | |
| | | Rehabilitation code | 0 | 0.5190 | 0.5190 | O | O | | | | Prescription type code | 0 | 0.0013 | 0.0013 | | | |
| 12 | Results of pathology readings | Gender | 0 | 0.0359 | 0.0359 | | | | | | Prescription code | 0 | 0.0288 | 0.0288 | | | |
| | | Date of birth | 0.0014 | 0.7914 | 0.7928 | O | O | O | 18 | Diagnostic information | Date of birth | 0.0041 | 0.7602 | 0.7643 | O | O | O |
| | | Age at the examination | 0 | 0.3000 | 0.3000 | O | | | | | Gender | 0 | 0.0100 | 0.0100 | | | |
| | | Inspection classification code | 0 | 0.0010 | 0.0010 | | | | | | Age of diagnosis | 0 | 0.0744 | 0.0744 | | | |
| | | Main sampling site | 0.0433 | 0.3096 | 0.3529 | O | | | | | Clinical diagnosis(ICD-10th code) | 0.0054 | 0 | 0.0054 | | | |
| 13 | Surgical prescription details | Gender | 0 | 0.0011 | 0.0011 | | | | | | Disease code(ICD-10th) | 0.0028 | 0 | 0.0028 | | | |
| | | Date of birth | 0.0197 | 0.7075 | 0.7272 | O | O | | | | Disease name(English,ICD-10th) | 0.0021 | 0 | 0.0021 | | | |
| | | Age at the prescription | 0 | 0.0490 | 0.0490 | | | | | | Disease name(Korea,ICD-10th) | 0.0019 | 0 | 0.0019 | | | |
| | | Operation code | 0.0048 | 0.0403 | 0.0451 | | | | 19 | Visit information | Date of birth | 0.0003 | 0.9995 | 0.9998 | O | O | O |
| | | Operation name(English) | 0.0025 | 0 | 0.0025 | | | | | | Gender | 0 | 0.3272 | 0.3272 | O | | |
| | | Operation name(Korea) | 0.0003 | 0 | 0.0003 | | | | | | Age of arrival | 0 | 0.8777 | 0.8777 | O | O | O |

## 4. Discussion

The present study conducted an experiment based on the method of selecting objective QIs for CRDW. Even though 137 clinical data set columns were selected (Table 5) based on the QI selection method of the HIPAA rules, the present investigation was able to reduce the range of columns to 17 - 28 by using the objective QI selection method proposed in the present study. If the current QI selection method is used, more than 100 columns are selected as QIs, processed for de-identification, and consequently data utility drops. However, data utility may improve if limited number of columns that have re-identifiability are selected using uniqueness and influence values and the columns that do not have re-identifiability are removed.

A flexible de-identification method was prepared by indexing and scoring information such as the re-identification capability of data users and their affiliated institutions. Based on the scores obtained, stricter selection of QIs was applied for users who had higher de-identifiability while selecting less number of QIs for the users whose de-identifiability was low despite using identical data.

Even though the de-identification intent of users and institutions was objectively scored, there is, however, the possibility of users with low scores and less amount of de-identified data misusing information compared to the users with high scores if they have the intention to do so.[28] Therefore, an honest broker is needed for using the objective QI selection method proposed in the present study. The users who were graded using the method proposed for assessing user's re-identification intent must undergo not only the intervention by the honest broker but also check the usage log of the provided data. Furthermore, there are occasions wherein users need data with minimum level de-identification depending on the research purpose. The honest broker should flexibly provide the level of de-identification by relatively decreasing re-identifiability by strengthening the de-identification measures of other QIs.

The columns of clinical data set have diverse information, and application of the proposed selection method to all research data is limited. Accordingly, the present study proposed a method for extracting QIs from data and selecting QIs to be minimally used based on clinical data sets. The study offered a method to select QIs based on objective grounds rather than the experience and subjective judgments of researchers by preparing objective indices for selecting the final QIs by numerically calculating the uniqueness and influence of the selected QI data. Through the method, it is expected that researchers can not only protect research data from data linkage attacks that identify specific individuals by linking QIs with external data but also prevent disturbances to research by minimizing the data loss that occurs in the process of de-identification. The de-identification method for the finally selected QIs should make the unnecessary information in the information searched for research purpose unsearchable. In the case of information needed for research purpose, however, the information should be excluded from de-identification for use while strengthening de-identification of other data. In future research, an in-depth investigation can focus on the existing de-identification method, security, and efficiency to strengthen it further through differentiating the level of de-identification by weighting the finally selected QIs in the present study.

---

**Summary points**

· With the vast amounts of data handled in clinical settings, de-identifying data appropriately in a manner that retains data utility while also protecting personal information is essential.

· The present study proposed a method for extracting Quasi Identifiers (QIs) from data and selecting QIs to be minimally used based on clinical data sets.

> · The study offered a method to select QIs based on objective grounds rather than the experience and subjective judgments of researchers by preparing objective indices for selecting the final QIs by numerically calculating the uniqueness and influence of the selected QI data.
> · The proposed methodology can minimize the danger of information leakage when used for research by identifying connections in data based on the risk of exposing information when data is collated and minimize data loss through de-identification processes.

# 5. Conflict of Interest

No potential conflict of interest relevant to this article was reported.

# 6. Funding

This study was supported by a grant from the National R&D Program for Cancer Control, Ministry of Health and Welfare, Republic of Korea (1631180).

# 7. References

[1] F. Prasser, F. Kohlmayer, K.A. Kuhn, Efficient and effective pruning strategies for health data de-identification, BMC Med Inform Decis Mak, 16 (2016) 49.
[2] A. Narayanan, An adversarial analysis of the reidentifiability of the heritage health prize dataset, (2011).
[3] S.Y. Shin, Y. Lyu, Y. Shin, H.J. Choi, J. Park, W.S. Kim, J.H. Lee, Lessons Learned from Development of De-identification System for Biomedical Research in a Korean Tertiary Hospital, Healthc Inform Res, 19 (2013) 102-109.
[4] U.D.o. Health, H. Services, Guidance regarding methods for de-identification of protected health information in accordance with the Health Insurance Portability and Accountability Act (HIPAA) Privacy Rule, 2015.
[5] Health Insurance Portability and Accountability Act of 1996 Safe Harbor Method.
[6] Personal information de-identification management guideline, in: O.f.G.P. Coordination (Ed.), 2016.
[7] S.L. Garfinkel, De-identification of personal information, National Institute of Standards and Technology, Gaithersburg, MD, Tech. Rep. IR-8053, (2015).
[8] K. El Emam, F.K. Dankar, R. Issa, E. Jonker, D. Amyot, E. Cogo, J.-P. Corriveau, M. Walker, S. Chowdhury, R. Vaillancourt, A globally optimal k-anonymity method for the de-identification of health data, Journal of the American Medical Informatics Association, 16 (2009) 670-682.

[9] M.M. Merener, Theoretical results on de-anonymization via linkage attacks, Transactions on Data Privacy, 5 (2012) 377-402.

[10] Y.J. Lee, K.H. Lee, Re-identification of medical records by optimum quasi-identifiers, Advanced Communication Technology (ICACT), 2017 19th International Conference on, IEEE, 2017, pp. 428-435.

[11] A.M. Omer, M.M.B. Mohamad, SIMPLE AND EFFECTIVE METHOD FOR SELECTING QUASI-IDENTIFIER, Journal of Theoretical and Applied Information Technology, 89 (2016) 512.

[12] C. Graham, Anonymisation: managing data protection risk code of practice, Information Commissioner's Office, (2012).

[13] K. El Emam, L. Arbuckle, G. Koru, B. Eze, L. Gaudette, E. Neri, S. Rose, J. Howard, J. Gluck, De-identification methods for open health data: the case of the Heritage Health Prize claims dataset, J Med Internet Res, 14 (2012) e33.

[14] Y.-R. Lee, Y.-C. Chung, J.-S. Kim, H.-K. Park, Personal Health Information De-identified Performing Methods in Big Data Environments, International Journal of Software Engineering and Its Applications, 10 (2016) 127-138.

[15] S. Kim, H. Lee, Y.D. Chung, Privacy-preserving data cube for electronic medical records: An experimental evaluation, Int J Med Inform, 97 (2017) 33-42.

[16] M. Scaiano, G. Middleton, L. Arbuckle, V. Kolhatkar, L. Peyton, M. Dowling, D.S. Gipson, K.E. Emam, A Unified Framework for Evaluating the Risk of Re-identification of Text De-identification Tools, J Biomed Inform, (2016).

[17] A. Dehghan, A. Kovacevic, G. Karystianis, J.A. Keane, G. Nenadic, Combining knowledge- and data-driven methods for de-identification of clinical narratives, J Biomed Inform, 58 Suppl (2015) S53-59.

[18] M. Li, D. Carrell, J. Aberdeen, L. Hirschman, B.A. Malin, De-identification of clinical narratives through writing complexity measures, Int J Med Inform, 83 (2014) 750-767.

[19] C. Grouin, A. Neveol, De-identification of clinical notes in French: towards a protocol for reference corpus development, J Biomed Inform, 50 (2014) 151-161.

[20] G. Zuccon, D. Kotzur, A. Nguyen, A. Bergheim, De-identification of health records using Anonym: effectiveness and robustness across datasets, Artif Intell Med, 61 (2014) 145-151.

[21] A.C. Fernandes, D. Cloete, M.T. Broadbent, R.D. Hayes, C.K. Chang, R.G. Jackson, A. Roberts, J. Tsang, M. Soncul, J. Liebscher, R. Stewart, F. Callard, Development and evaluation of a de-identification procedure for a case register sourced from mental health electronic records, BMC Med Inform Decis Mak, 13 (2013) 71.

[22] K.Y. Aryanto, G. van Kernebeek, B. Berendsen, M. Oudkerk, P.M. van Ooijen, Image De-Identification Methods for Clinical Research in the XDS Environment, J Med Syst, 40 (2016) 83.

[23] S.Y. Shin, Y.R. Park, Y. Shin, H.J. Choi, J. Park, Y. Lyu, M.S. Lee, C.M. Choi, W.S. Kim, J.H. Lee, A De-identification method for bilingual clinical texts of various note types, J Korean Med Sci, 30 (2015) 7-15.

[24] C.A. Kushida, D.A. Nichols, R. Jadrnicek, R. Miller, J.K. Walsh, K. Griffin, Strategies for de-

identification and anonymization of electronic health record data for use in multicenter research studies, Med Care, 50 Suppl (2012) S82-101.

[25] S.M. Meystre, O. Ferrandez, F.J. Friedlin, B.R. South, S. Shen, M.H. Samore, Text de-identification for privacy protection: a study of its impact on clinical text information content, J Biomed Inform, 50 (2014) 142-150.

[26] V.N. Patel, D.C. Kaelber, Using aggregated, de-identified electronic health record data for multivariate pharmacosurveillance: a case study of azathioprine, J Biomed Inform, 52 (2014) 36-42.

[27] F.K. Dankar, K. El Emam, A. Neisa, T. Roffey, Estimating the re-identification risk of clinical data sets, BMC Med Inform Decis Mak, 12 (2012) 66.

[28] H.J. Choi, M.J. Lee, C.M. Choi, J. Lee, S.Y. Shin, Y. Lyu, Y.R. Park, S. Yoo, Establishing the role of honest broker: bridging the gap between protecting personal health data and clinical research efficiency, PeerJ, 3 (2015) e1506.